\documentclass{amsart}
\usepackage{amsaddr}

\usepackage{bbold}
\DeclareSymbolFont{bbold}{U}{bbold}{m}{n}
\DeclareSymbolFontAlphabet{\mathbbold}{bbold}

\usepackage{txfonts}

\usepackage{graphicx}

\begin{document}

\title[Bosonic Dynamical Mean-Field Theory]{Bosonic Dynamical Mean-Field Theory }



\author[M. Snoek]{Michiel Snoek}
\address{Institute for Theoretical Physics, University of Amsterdam, 1090 GL Amsterdam, The Netherlands 
}


\author[W. Hofstetter]{Walter Hofstetter}
\address{Institut f\"ur Theoretische Physik, Johann Wolfgang Goethe-Universit\"at, 60438 Frankfurt/Main, Germany
}

\begin{abstract}
We derive the Bosonic Dynamical Mean-Field equations for bosonic atoms in optical lattices with arbitrary lattice geometry. The equations are presented as a systematic expansion in $1/z$, $z$ being the number of lattice neighbours. Hence the theory is applicable in sufficiently high dimensional lattices. We apply the method to a two-component mixture, for which a rich phase diagram with spin-order is revealed.\\
\end{abstract}

\maketitle

\section{Introduction}

Bosonic atoms in optical lattices, described by the Bose-Hubbard Hamiltonian \cite{fisher_weichman_89, jaksch_bruder_98}, can easily be brought into the strongly interacting regime by increasing the intensity of the laser beams forming the optical lattice \cite{bloch_dalibard_08}. 
At commensurate fillings this is reflected in the formation of a Mott Insulating state. This transition has been detected experimentally \cite{greiner_esslinger_02}, as one of the first quantum phase transitions in the strongly correlated regime in the field of cold atoms.

In the case of a mixture of different bosonic components, either of different hyperfine states of the same atom or different species, additional spin order can exist in the Mott insulating state, thus further enriching the phase diagram \cite{kuklov_svistunov_03, duan_demler_03, altman_hofstetter_03, isacsson_cha_05, powell_09, soyler_capogrosso_09, hubener_snoek_09, capogrosso_soyler_10}.

Whereas to describe the single-species Mott insulator transition it is sufficient to use Gutzwiller Mean-Field Theory \cite{rokhsar_kotliar_91, sheshadri_krishnamurthy_93}, which uses the local superfluid order parameter as the mean field, to capture spin order one needs more sophisticated methods. The reason is that the superfluid order parameter vanishes in the insulating states, effectively decoupling the sites, such that all possibilities for spin-order are degenerate. Here we formulate Bosonic Dynamical Mean-Field Theory (BDMFT) \cite{byczuk_vollhardt_08, hubener_snoek_09, hu_tong_09, anders_gull_10, byczuk_vollhardt_10, byczuk_vollhardt_10b, anders_gull_11}, which is able to describe long-range spin order, because in addition to the Gutzwiller Mean-Field term describing the superfluid-insulator transition, also a dynamical coupling is present, essential for the description of virtual hopping processes in the insulating states. The theory is presented here as a systematic expansion in $1/z$, $z$ being the lattice coordination number.   

A generic bosonic mixture consisting of $N$ flavors in a sufficiently deep optical lattice is described by the following Bose-Hubbard Hamiltonian
\begin{equation}
\hat{\mathcal{H}} = -  \sum_{\langle i j \rangle, \alpha} J_{\alpha} \left( \hat b_{i\alpha}^\dagger \hat b_{j\alpha}^{\phantom\dagger} + {\rm h.c.} \right) 
 - \sum_{i, \alpha} \mu_\alpha  \hat n_{i \alpha} 
 + \frac{1}{2} \sum_{i, \alpha, \gamma} U_{\alpha \gamma} \hat n_{i \alpha} \left(\hat n_{i \gamma} - \delta_{\alpha \gamma} \right)
 \label{snoek_bh}
\end{equation}
Here Greek indices ($\alpha, \gamma$) denote the flavors and Roman indices ($i,j$) denote the lattice sites, such that $\hat b_{i\alpha}^{(\dagger)} $ denotes the annihilation (creation) operator of a boson with flavor $\alpha$ at site $i$. We have introduced the number operator $\hat n_{i \alpha} = \hat b_{i\alpha}^{\dagger}\hat b_{i\alpha}^{\phantom\dagger}$. The tunneling parameters $J_{\alpha}$ can be different for different species, which is relevant when the atoms have a different mass or a different coupling to the laser beams. Also the chemical potentials $\mu_\alpha$ can be different, 
in order to allow independent tuning of the densities of the different components. The inter- and intraspecies interactions are encoded in $U_{\alpha \gamma}$.

\section{Methodology} 

In deriving the BDMFT equations we consider the limit of a high- (but finite-) dimensional optical lattice \cite{hubener_snoek_09, georges_kotliar_96}. Hence $z$, the number of neighbouring lattice sites, is large and $1/z$ can serve as an expansion parameter. In order for the total energy per site to converge in the limit $z \to \infty$, the tunnelling amplitudes then need to be rescaled as $J_{\alpha} = J_{\alpha}'/z$. This introduces the small control parameter $1/z$ into the Hamiltonian.  At the final step in the derivation we will return from $J_{\alpha}'$ to $J_{\alpha}$. We thus present the BDMFT equations following a systematic expansion in the small parameter $1/z$  \cite{hubener_snoek_09}. This is different from the original proposal \cite{byczuk_vollhardt_08, byczuk_vollhardt_10, byczuk_vollhardt_10b}, as well as from a recent derivation \cite{anders_gull_10, anders_gull_11}, although the final equations in all cases (for finite dimensions) coincide.

\subsection{Effective Action}

Using the conventional coherent state representation (see e.g. \cite{stoof_gubbels_book_09}),
the partition function $Z$ is expressed as
\begin{equation}
Z = \int D[\mathbf{b}^*] D[\mathbf{b}] \exp \left( - S[\mathbf{b}^*, \mathbf{b}]/\hbar \right),
\end{equation}
where we use the shorthand notation $\int D[\mathbf{b}^*] D[\mathbf{b}] = \int \prod_{i,\alpha} D[b_{i\alpha}^*] D[b_{i\alpha}]$, and the $b_{i\alpha}(\tau)$ are complex-valued fields.\footnote{
This is necessarily a functional integral, as we must account for a continuum of paths taken by the $b_{i\alpha}$ (and their complex conjugates) as functions of the continuous variable $\tau$. Also note, in keeping with convention, we refer to the discrete field amplitudes $b_{i\alpha}$ simply as fields.}
The action $S[\mathbf{b}^*, \mathbf{b}]$ is given by
\begin{multline}
S[\mathbf{b}^*, \mathbf{b}]  =  \int_0^{\hbar\beta} d \tau \left\{ \sum_{i,\alpha} b_{i\alpha}^*(\tau) \left(\hbar \frac{\partial}{\partial\tau} - \mu_\alpha \right) b_{i\alpha}(\tau)
 - \sum_{\langle i j \rangle, \alpha} \frac{J_{\alpha}'}{z} \left[ b_{i\alpha}^*(\tau)b_{j\alpha}(\tau) + {\rm c.c.} \right] 
\right. \\ \left. 
+ \frac{1}{2} \sum_{i, \alpha, \gamma} U_{\alpha \gamma} b_{i \alpha}^*(\tau)b_{i \alpha}(\tau) \left[ b_{i \gamma}^*(\tau)b_{i \gamma}(\tau) - \delta_{\alpha \gamma} \right] \right\}.
\end{multline}

Following the same `cavity' method used to derive the fermionic DMFT equations \cite{georges_kotliar_96}, we now consider a specific site and call it site $0$. We split the action into three parts: $S_0$ contains those terms exclusively related to site $0$, $S^{(0)}$ contains those terms not incorporating site $0$, and $\Delta S$ contains the terms connecting site $0$ to the other sites. The system with site $0$ excluded is called the cavity system.  $\Delta S$ is thus given by
\begin{equation}
\Delta S[\mathbf{b}^*, \mathbf{b}] = - \frac{1}{z} \int_0^{\hbar\beta} d \tau  \sum_{\langle 0 j \rangle, \alpha} J_{\alpha}' \left[ b_{0\alpha}^*(\tau)b_{j\alpha}(\tau) + {\rm c.c.} \right], 
\end{equation}
which is clearly proportional to the expansion parameter $1/z$. We exploit this property to systematically expand the action up to second order in $\Delta S$, in order to derive an effective action $S_{\rm eff}^0$ for site $0$.  The effective action is defined through the corresponding partition function $Z_{\rm eff}^0 \equiv Z/Z^{(0)}$, where
\begin{align}
Z_{\rm eff}^0 =& \int D[\mathbf{b}_0^*] D[\mathbf{b}_0] \exp \left( - S_{\rm eff}^0[\mathbf{b}_0^*, \mathbf{b}_0]/\hbar \right),
\label{Eq:HofstetterCavityZEff}
\\
Z^{(0)} =& \int D^{(0)}[\mathbf{b}^*] D^{(0)}[\mathbf{b}] \exp \left( - S^{(0)}[\mathbf{b}^*, \mathbf{b}]/\hbar \right).
\label{Eq:HofstetterCavityZ}
\end{align}
In Eq.\ \eqref{Eq:HofstetterCavityZ} the integral specifically excludes the site $0$ fields , as indicated by the notation $D^{(0)}$.  This is equivalent to performing all integrals appearing in the definition of $Z$, except those located at site $0$. In this way we obtain
\begin{equation}
\begin{split}
Z_{\rm eff}^0 =& \frac{1}{Z^{(0)}} \int D[\mathbf{b}_0^*] D[\mathbf{b}_0] \exp \left( - S_0[\mathbf{b}_0^*, \mathbf{b}_0]/\hbar \right)  
\\ & \times 
\int D^{(0)}[\mathbf{b}^*] D^{(0)}[\mathbf{b}] \exp \left( - \left\{ S^{(0)}[\mathbf{b}^*, \mathbf{b}] + \Delta S[\mathbf{b}^*, \mathbf{b}]  \right \} /\hbar \right),
\end{split}
\end{equation}
which can be expanded in powers of $\Delta S$ through
\begin{multline}
\int D^{(0)}[\mathbf{b}^*] D^{(0)}[\mathbf{b}]   
\exp \left( - \left\{ S^{(0)}[\mathbf{b}^*, \mathbf{b}] + \Delta S[\mathbf{b}^*, \mathbf{b}]  \right \} /\hbar \right) \\ =
\sum_{k=0}^\infty \frac{1}{\hbar^{k} k!} \int D^{(0)}[\mathbf{b}^*] D^{(0)}[\mathbf{b}]  (-\Delta S[\mathbf{b}^*, \mathbf{b}])^k   
\exp \left( - S_0[\mathbf{b}_0^*, \mathbf{b}_0]/\hbar \right).
\end{multline}
Using now the definition for an expectation value in the cavity system 
$$\langle A \rangle_{(0)} = (Z^{(0)})^{-1} \int D^{(0)}[\mathbf{b}^*] D^{(0)}[\mathbf{b}]  A[\mathbf{b}^*, \mathbf{b}]   e^{ - S^{(0)}/\hbar},$$
this immediately gives us
\begin{equation}
Z_{\rm eff}^0 = \int D[\mathbf{b}_0^*] D[\mathbf{b}_0] \exp \left( - S_0[\mathbf{b}_0^*, \mathbf{b}_0]/\hbar \right) \sum_{k=0}^{\infty} \frac{\langle(-\Delta S)^k \rangle_{(0)}}{\hbar^k k!}.
\label{Eq:HofstetterZEffPower}
\end{equation}
Note that the terms $\langle(\Delta S)^k \rangle_{(0)}$ still depend on the fields at site $0$, and cannot be taken outside the integral.  We determine explicitly the two lowest orders in $\Delta S$, i.e., to the order in $\Delta S$ we wish to consider:
\begin{align} 
\langle \Delta S \rangle_{(0)}
=&-\frac{1}{z} \int_0^{\hbar\beta} d \tau  \sum_{\langle 0 j \rangle, \alpha} J_{\alpha}' \left[ b_{0\alpha}^* (\tau) \langle b_{j\alpha} (\tau) \rangle_{(0)} + {\rm c.c.} \right], 
\label{Eq:HofstetterDeltaS}
\\
\begin{split}
\langle (\Delta S)^2 \rangle_{(0)} 
=&\frac{1}{z^2} \iint_0^{\hbar\beta} d \tau d \tau'  \sum_{\langle 0 j \rangle, \alpha} \sum_{\langle 0 j' \rangle, \gamma} J_{\alpha}'  J_{\gamma}' 
\Big[ 
 b_{0\alpha}^* (\tau)  b_{0\gamma}^* (\tau') \langle b_{j\alpha} (\tau) b_{j'\gamma} (\tau')  \rangle_{(0)} 
\\ 
&+ b_{0\alpha}^* (\tau)  b_{0\gamma} (\tau') \langle b_{j\alpha} (\tau) b_{j'\gamma}^* (\tau')  \rangle_{(0)} 
+ \textrm{c.c.}
\Big]. 
\end{split}
\end{align}
The final (re-exponentiation) step is to note that, taking the effective action to be
\begin{equation}
S_{\rm eff}^0[\mathbf{b}_0^*, \mathbf{b}_0] = S_0[\mathbf{b}_0^*, \mathbf{b}_0] + \langle \Delta S \rangle_{(0)} - \frac{1}{2 \hbar} \left[ \langle (\Delta S)^2 \rangle_{(0)} - \langle \Delta S \rangle_{(0)}^2\right],
\end{equation}
it follows that taking $\exp(-S_{\rm eff}^0[\mathbf{b}_0^*, \mathbf{b}_0]/\hbar)$ to second order in $\Delta S$ within Eq.~\eqref{Eq:HofstetterCavityZEff} yields Eq.~\eqref{Eq:HofstetterZEffPower} to second order in $\Delta S$ (and consequently in $1/z$), as desired.  The second-order terms can be conveniently expressed in terms of the connected Green's function in the cavity system, through 
$\langle \Delta S \rangle_{(0)}^2  - \langle (\Delta S)^2 \rangle_{(0)} = 
\iint_0^{\hbar\beta} d \tau d \tau'  \sum_{\langle 0 j \rangle, \alpha} \sum_{\langle 0 j' \rangle, \gamma} J_{\alpha}'  J_{\gamma}' 
[\mathbf{B}_{\alpha}^{*\mathrm{T}}(\tau)
\mathbf{G}_{j\alpha; j'\gamma}^{(0)} (\tau, \tau')
\mathbf{B}_{\gamma}(\tau')]$, where $\mathbf{B}_{\alpha}^{*\mathrm{T}}(\tau) = (b_{0\alpha}^* (\tau) , b_{0\alpha} (\tau))$ and
\begin{equation}
\mathbf{G}_{j\alpha; j'\gamma}^{(0)} (\tau, \tau') = 
\left(\begin{array}{cc} 
G_{j\alpha; j'\gamma}^{(0)d} (\tau, \tau') & G_{j\alpha; j'\gamma}^{(0)o} (\tau, \tau') 
\\
G_{j'\gamma; j\alpha}^{(0)o*} (\tau', \tau) & G_{j'\gamma; j\alpha}^{(0)d} (\tau', \tau) 
\end{array}
 \right).
 \label{Eq:HofstterGreensCavity}
\end{equation}
In the above, we have
$G_{j\alpha; j'\gamma}^{(0)d} (\tau, \tau') =  \langle b_{j\alpha} (\tau)\rangle_{(0)}\langle b_{j'\gamma}^* (\tau')  \rangle_{(0)} - \langle b_{j\alpha} (\tau) b_{j'\gamma}^* (\tau')  \rangle_{(0)}$, and
$
G_{j\alpha; j'\gamma}^{(0)o} (\tau, \tau')   =  \langle b_{j\alpha} (\tau)  \rangle_{(0)} \langle b_{j'\gamma} (\tau')  \rangle_{(0)}  - \langle b_{j\alpha} (\tau) b_{j'\gamma} (\tau')  \rangle_{(0)} 
$.  Note that the Green's function has a matrix form in spin-space as well as Nambu (particle-hole) space; the latter is due to possible off-diagonal long-range superfluid order.

Since the action is invariant under imaginary-time translations single particle expectations values like $\langle b_{j \alpha} (\tau) \rangle $ do not depend on $\tau$. Correlation functions  only depend on the imaginary-time difference $\tau-\tau'$. This latter fact is used in the following to write the equations in terms of the bosonic Matsubara frequencies $\omega_n = 2 \pi n/\hbar \beta$, reciprocal to this imaginary-time difference.  

\subsection{Self-Consistency Relations}
To obtain a closed self-consistency loop, expectation values in the cavity system (i.e., $\langle O \rangle_{(0)}$), must be identified with expectation values on the impurity site (i.e., $\langle O \rangle_{0}$). In the case of the BDMFT equations, this step requires a careful treatment, as possible $1/z$ corrections must be taken into account.

We first consider the second-order terms.  Because terms to second order in $\Delta S$  within the effective action already appear at second order in $1/z$, any $1/z$ corrections lead to an irrelevant contribution and we need consider leading order behaviour only.  We can therefore apply similar reasoning as for the fermionic case \cite{georges_kotliar_96}, and take the limit of infinite dimensions in dealing with these terms.

We combine all terms within the effective action quadratic in the impurity site fields into the Weiss field: 
\begin{equation}
\boldsymbol{\mathcal{G}}_{\alpha \gamma}^{-1}(i \omega_n) = 
\delta_{\alpha \gamma} (i \hbar \omega_n \sigma_z + \mu_\alpha  \mathbf{I}_2) - \sum_{\langle 0i \rangle, \langle 0j \rangle} \frac{t'_\alpha J'_{\gamma}}{z^2} \mathbf{G}_{i\alpha; j\gamma}^{(0)} (i \omega_n), 
\end{equation}
where $\mathbf{I}_{2}$ is the $2\times 2$ identity matrix. From the fact that the self-energy is a local (i.e., momentum-independent) quantity in infinite dimensions \cite{georges_kotliar_96}, 
then follows the local Dyson equation
\begin{equation}
\boldsymbol{\mathcal{G}}_{\alpha \gamma}^{-1}(i \omega_n) = \mathbf{G}_{\alpha \gamma}^{-1}(i \omega_n) + \boldsymbol{\Sigma}_{\alpha \gamma}(i \omega_n),
\end{equation}
For future use we also define the hybridization function 
$
\boldsymbol{\Delta}_{\alpha \gamma} (i \omega_n) = \delta_{\alpha \gamma}(i \hbar \omega_n \sigma_z + \mu_\alpha   \mathbf{I}_2) - \mathbf{G}_{\alpha \gamma}^{-1}(i \omega_n) - \boldsymbol{\Sigma}_{\alpha \gamma}(i \omega_n) 
$,
such that 
$
\boldsymbol{\mathcal{G}}_{\alpha \gamma}^{-1}(i \omega_n) = \delta_{\alpha \gamma} (i \hbar \omega_n \sigma_z + \mu_\alpha \mathbf{I}_2 )-
\boldsymbol{\Delta}_{\alpha \gamma} (i \omega_n)
$,
and also 
\begin{equation}
\boldsymbol{\Delta}_{\alpha \gamma} (i \omega_n) =  \sum_{\langle 0j \rangle, \langle 0j' \rangle} \frac{t'_\alpha J'_{\gamma}}{z^2} \mathbf{G}_{j\alpha; j'\gamma}^{(0)} (i \omega_n) .
\end{equation}

The superfluid order parameter $\langle \hat b_\alpha \rangle$, on the other hand, appears at first order in $1/z$ in the effective action. To be consistent, we must therefore take possible $1/z$ corrections into account, which occur because we must calculate the expectation value of $\hat b_\alpha$ in the cavity system. We now show that this indeed leads to an important correction. Qualitatively this can be understood as being because the sites $j$ on which $\langle \hat b_\alpha \rangle$ is calculated have one fewer neighbour in the cavity system, as the impurity site has been removed. Previously this motivated us to implement the $1/z$ correction by first order perturbation theory in the missing neighbour, which on the Bethe lattice gives rise to results for the superfluid-insulator transition very close to the numerically exact solution \cite{hubener_snoek_09, semerjian_tarzia_09}. Here we show that we can also derive a closed expression without invoking perturbation theory.

First of all we remark that in the original homogeneous system the expectation value is independent of the lattice position and in particular is equal to that at site $0$ (chosen as the impurity site): $\langle b_{i\alpha \rangle} = \langle b_{0\alpha} \rangle$. Secondly, expectation values on the impurity site are calculated including all relevant orders in $1/z$ and thus constitute the DMFT-approximation for the expectation values in the original system: $\langle O_0 \rangle_0 = \langle O_0 \rangle$.
For the expectation values in the cavity system this is different, because the impurity site is missing.

To calculate the expectation value $b_{j\alpha}$ for a site in the cavity system we couple the bosonic fields at site $j$ to a generic source term $\mathcal{J}_{j\alpha}$ \cite{stoof_gubbels_book_09}. Hence, the partition function
\begin{equation}
Z[\mathcal{J}_{j\alpha}, \mathcal{J}_{j\alpha}^*] = \int D[\mathbf{b}^*] D[\mathbf{b}] 
e^{- S[\mathbf{b}^*, \mathbf{b}]/\hbar + \int d \tau [b_{j \alpha}^*(\tau) \mathcal{J}_{j \alpha}(\tau) + \mathcal{J}_{j\alpha}^*(\tau) b_{j \alpha}(\tau)]}
\end{equation}    
depends on this source, and we obtain the expectation value from the functional derivative 
\begin{equation}
\langle b_{j\alpha} (\tau) \rangle = \frac{\delta}{\delta \mathcal{J}_{j \alpha}^*} \ln(Z[\mathcal{J}_{j\alpha}, \mathcal{J}_{j\alpha}^*]).
\label{Eq:HofstetterExpectationDef}
\end{equation}
We now carry out the cavity construction with the sources present in order to derive $Z_{\rm eff}^{0} [\mathcal{J}_{j\alpha}, \mathcal{J}_{j\alpha}^*]$.  Since we are only interested in calculating the superfluid order parameter, we only keep terms linear in the sources; to calculate $1/z$ corrections to the Green's function one must retain second order terms as well.

Defining the shorthand $S_{J} =  \int d \tau [b_{j \alpha}^*(\tau) \mathcal{J}_{j \alpha}(\tau) + \mathcal{J}_{j\alpha}^*(\tau) b_{j \alpha}(\tau) ]$, we obtain:
\begin{equation}
\begin{split}
Z_{\rm eff}^0 [\mathcal{J}_{j\alpha}, \mathcal{J}_{j\alpha}^*] = & \frac{1}{Z^{(0)}} \int D[\mathbf{b}^*] D[\mathbf{b}] e^{ - S[\mathbf{b}^*, \mathbf{b}]/\hbar + S_J } 
\\
= & \frac{1}{Z^{(0)}} \int D[\mathbf{b}^*] D[\mathbf{b}] e^{ -(S_0[\mathbf{b}_{0}^*, \mathbf{b}_{0}] + S^{(0)}[\mathbf{b}^*, \mathbf{b}])/\hbar}
\\ & \times 
\left[ 1 - \frac{\Delta S}{\hbar} 
+ \frac{1}{2\hbar^2} (\Delta S)^2 + S_J - \frac{S_J \Delta S}{\hbar} + \cdots  \right].
\end{split}
\end{equation}
As previously, we then integrate out the fields in the cavity system and re-exponentiate, yielding
\begin{equation}
\begin{split}
Z_{\rm eff}^0 [\mathcal{J}_{j\alpha}, \mathcal{J}_{j\alpha}^*] =& \int D[\mathbf{b}_0^*] D[\mathbf{b}_0] e^{ - S_0[\mathbf{b}_0^*, \mathbf{b}_0]/\hbar}
\left[ 1 - \frac{\langle \Delta S \rangle_{(0)}}{\hbar} 
\right. \\ & \left.
+ \frac{1}{2 \hbar^2} \langle (\Delta S)^2 \rangle_{(0)} + \langle S_J\rangle_{(0)} - \frac{\langle S_J \Delta S\rangle_{(0)}}{\hbar} + \cdots  \right] \\ 
=& \int D[\mathbf{b}_0^*] D[\mathbf{b}_0] e^{-S_{\rm eff}^0[\mathbf{b}_0^*, \mathbf{b}_0]/\hbar +  \langle S_J \rangle_{(0)} -  (\langle S_J \Delta S \rangle_{(0)} -\langle S_J \rangle_{(0)} \langle \Delta S \rangle_{(0)})/\hbar },
\end{split}
\label{Eq:HofstetterZeffJ}
\end{equation}
for which we require the following expectation values:
\begin{align}
\langle S_J \rangle_{(0)}  =& \int_{0}^{\hbar\beta} d \tau \left[
\langle b_{j\alpha} (\tau) \rangle_{(0)}^* \mathcal{J}_{j\alpha} (\tau) + \mathcal{J}_{j\alpha}^* (\tau) \langle b_{j\alpha} (\tau) \rangle_{(0)} 
\right],
\\
\begin{split}
\langle S_J \Delta S \rangle_{(0)}  =&  - \sum_{\langle 0 j' \rangle, \gamma} \frac{J'_{\gamma}}{z} \iint_{0}^{\hbar\beta} d \tau d \tau' \Bigg\{ 
\mathcal{J}_{j\alpha}^*(\tau) \Big[ b_{0 \gamma}^* (\tau') \langle b_{j'\gamma} (\tau') b_{j \alpha} (\tau) \rangle_{(0)} 
 \\ & 
 + b_{0 \gamma} (\tau') \langle b_{j'\gamma}^* (\tau') b_{j \alpha}(\tau) \rangle_{(0)} \Big]  
 + \textrm{c.c}\Bigg\}.
\end{split}
\end{align}
Substituting these together with Eq.~\eqref{Eq:HofstetterDeltaS} into Eq.~\eqref{Eq:HofstetterZeffJ}, we then use Eq.\ \eqref{Eq:HofstetterExpectationDef}, and determine the following result for the superfluid order parameter:
\begin{equation}
\begin{split}
\langle b_{j\alpha} (\tau) \rangle 
=& \langle b_{j\alpha} (\tau) \rangle_{(0)} + \frac{1}{\hbar} \sum_{\langle 0 j' \rangle, \gamma} \frac{J'_{\gamma}}{z} \int_{0}^{\hbar\beta}   d \tau' \Bigg \{ 
\\ &  
\times \langle b_{0 \gamma}^* (\tau') \rangle \left[ \langle b_{j'\gamma} (\tau') b_{j \alpha} (\tau) \rangle_{(0)} 
-  \langle b_{j'\gamma} (\tau') \rangle_{(0)} \langle b_{j \alpha} (\tau) \rangle_{(0)} \right]  
\\ & + 
 \langle b_{0 \gamma} (\tau') \rangle  \left[ \langle b_{j'\gamma}^* (\tau') b_{j \alpha}(\tau) \rangle_{(0)}
 - \langle b_{j'\gamma}^* (\tau') \rangle_{(0)} \langle b_{j \alpha}(\tau) \rangle_{(0)} \right]
 \Bigg\},
 \end{split}
 \end{equation}
 which can be expressed in terms of the Green's function [Eq.~\eqref{Eq:HofstterGreensCavity}] as:
 \begin{multline}
 \langle b_{j\alpha} (\tau) \rangle  = \langle b_{j\alpha} (\tau) \rangle_{(0)}
\\  - \frac{1}{\hbar} \sum_{\langle 0 j' \rangle, \gamma} \frac{J'_{\gamma}}{z} \int_{0}^{\hbar\beta}   d \tau' \Big[
   \langle b_{0 \gamma}^* (\tau') \rangle 
   G_{j\alpha; j'\gamma}^{(0)o} (\tau, \tau') +
  \langle b_{0 \gamma} (\tau') \rangle  
  G_{j\alpha; j'\gamma}^{(0)d} (\tau, \tau') \Big]
\end{multline}
Using now that the superfluid order parameter is independent of $\tau$, we can carry out the integral over $\tau'$ and obtain
\begin{equation}
\langle b_{j\alpha} \rangle = \langle b_{j\alpha} \rangle_{(0)} - \sum_{\langle 0 j' \rangle, \gamma} \frac{J'_{\gamma}}{z} \left[ 
   \langle b_{0 \gamma}^* \rangle G_{j\alpha; j'\gamma}^{(0)o} (\omega_n=0)+
  \langle b_{0 \gamma}  \rangle  G_{j\alpha; j'\gamma}^{(0)d} (\omega_n=0) \right],
\end{equation}
which indeed yields a $1/z$ correction to  $\langle b_{j\alpha} \rangle_{(0)}$ with respect to $\langle b_{j\alpha} \rangle$.
To complete this we note that that the source field coupling to $b_{0\alpha}^*$ in the effective action is given by 
$J_{\alpha} \sum_{\langle 0 j \rangle}  \langle b_{j\alpha} \rangle_{(0)}$.  This we can now re-express as
\begin{equation}
\begin{split}
J_{\alpha} \sum_{\langle 0 j\rangle}  \langle b_{j\alpha} \rangle_{(0)} =& 
J_{\alpha} \sum_{\langle 0 j\rangle} \left\{ \langle b_{j\alpha} \rangle + \sum_{\langle 0 j' \rangle, \gamma} \frac{J'_{\gamma}}{z} \left[ 
   \langle b_{0 \gamma}^* \rangle G_{j\alpha;j'\gamma}^{(0)o} (0)+
  \langle b_{0 \gamma}  \rangle  G_{j\alpha;j'\gamma}^{(0)d} (0) \right] \right\}  \\
  =& z J_{\alpha} \phi_\alpha +  \sum_{\langle 0 j'\rangle, \langle 0 j \rangle, \gamma} \phi_\gamma \frac{ J_{\alpha}' J_{\gamma}'}{z^{2}} \left\{ G_{j\alpha;j'\gamma}^{(0)o} (0) + G_{j\alpha;j'\gamma}^{(0)d} (0) \right\} \\
  =& z J_{\alpha} \phi_\alpha + \sum_\gamma \phi_\gamma \left [ \Delta_{\alpha \gamma}^d (0) +  \Delta_{\alpha \gamma}^o (0) \right]
\equiv z J_{\alpha} \phi_\alpha^{\rm in}, 
\end{split}
\label{snoek_eqA}
\end{equation}
where we have introduced $\phi_\alpha = \langle b_{j\alpha}\rangle$ (which is $j$-independent and assumed to be real) and expressed the answer in terms of diagonal and off-diagonal elements of the hybridization function at zero frequency. This constitutes the self-consistency equation for the superfluid order parameter.

For the special case of the Bethe lattice, the hybridization function is proportional to the Green's function. In this case, the expression derived from perturbation theory \cite{hubener_snoek_09} coincides with the closed expression in Eq.~\eqref{snoek_eqA}, explaining why the exact solution of the phase diagram for the single component bosons was so closely reproduced \cite{hubener_snoek_09}. 

\subsection{Implementation}
In order to complete a self-consistency loop, one has to extract the self-energy and the superfluid order parameter from the effective action such that a new input-superfluid order parameter and hybridization function can be calculated. We use the exact diagonalization method to do this, which means that the effective action is first mapped to a bosonic Anderson Hamiltonian. The Anderson Hamiltonian is not uniquely defined, and we make here the specific choice:  
\begin{equation}
\begin{split}
\hat{\mathcal{H}}_{\textrm{And}} =& - \sum_\alpha \mu_\alpha \hat n_\alpha + \frac{1}{2} \sum_{\alpha \gamma} U_{\alpha \gamma} \hat n_\alpha \left(\hat n_\gamma - \delta_{\alpha \gamma}\right) 
\\ &
+ \sum_{l} \left( \sum_\alpha \hat{\mathbf{b}}_\alpha^\dagger \mathbf{V}_{l \alpha} \hat{\mathbf{a}}_l + \hat{\mathbf{a}}_l^\dagger \boldsymbol{\epsilon}_{l} \hat{\mathbf{a}}_l \right) - z \sum_\alpha J_{\alpha} \left[ (\phi_\alpha^{\rm in})^* \hat b_\alpha + \hat b_\alpha^\dagger \phi_\alpha^{\rm in} \right].
\end{split}
\end{equation}
Here the $\hat b_\alpha^{(\dagger)}$ ($\hat n_\alpha =  \hat b_\alpha^\dagger \hat b_\alpha$) operators act on the impurity site. In addition we have introduced (annihilation) operators $\hat a_{l\alpha}^{(\dagger)}$ which act on orbitals labelled by $l$. These orbitals provide a non-interacting bath for the impurity site, which incorporates the hybridization with the other lattice sites. 
We have introduced a vector and matrix notation: 
$\hat{\mathbf{b}}_\alpha = (\hat{b}_\alpha, \hat{b}_\alpha^\dagger)^T$ and $\hat{\mathbf{a}}_l = (\hat{a}_l, \hat{a}_l^\dagger)^T$, and $\mathbf{V}_{l\alpha} = \left( \begin{smallmatrix} V_{l\alpha} & W_{l\alpha} \\ W_{l \alpha}^* & V_{l \alpha}^* \end{smallmatrix} \right)$ and $\boldsymbol{\epsilon}_{l} = \left( \begin{smallmatrix} \epsilon_l/2 & \delta_l \\ \delta_l^* & \epsilon_l/2 \end{smallmatrix} \right)$ are matrices in Nambu space. The  matrix $\boldsymbol{\epsilon}_{l}$ includes the on-site terms on orbital $l$ (note that there is no dependence on the spin index $\sigma$!), whereas $\boldsymbol{V}_{l\sigma}$ contains the diagonal and off-diagonal amplitudes for particles with spin index $\sigma$ to tunnel to orbital $l$.  
Integrating out the fields on the orbitals $a_l$, one obtains a single-site action with hybridization function:
\begin{equation}
\boldsymbol{\Delta}_{\alpha \gamma}^{\textrm{And}} (i \omega_n) = \frac{1}{4} \sum_l \mathbf{V}_{l \alpha}^* ( i \omega_n \sigma_z - \boldsymbol{\epsilon}_l)^{-1} \mathbf{V}_{l \gamma}.
\end{equation}
The parameters in the matrices $\mathbf{V}_{l\alpha}$ and $\boldsymbol{\epsilon}_l$ are obtained by fitting the Anderson hybridization function to the hybridization function obtained from the local Dyson equation. 
The Anderson Hamiltonian can then straightforwardly be implemented in the Fock basis. 
After diagonalization, the local Green's function can be obtained from the eigenstates and eigenenergies: in the Lehmann-representation we obtain
\begin{align}
G_{\alpha \gamma}^d (i \omega_n) =& \frac{1}{Z} \sum_{mn} \langle m | \hat b_\alpha | n\rangle \langle n | \hat b_\gamma^\dagger | m \rangle \frac{e^{- \beta E_n} - e^{-\beta E_m}}{E_n - E_m + i \hbar \omega_n} + \beta \phi_\alpha \phi_\gamma, \\
G_{\alpha \gamma}^o (i \omega_n) =& \frac{1}{Z} \sum_{mn} \langle m | \hat b_\alpha | n\rangle \langle n | \hat b_\gamma | m \rangle \frac{e^{- \beta E_n} - e^{-\beta E_m}}{E_n - E_m + i \hbar \omega_n} + \beta \phi_\alpha \phi_\gamma.
\end{align} 

Knowing the Green's function, the self-energy can in principle be determined from the local Dyson equation. However, this can lead to inaccurate results. A more accurate way is to use the equations of motion for the Green's function, in the same way as proposed for the fermionic case \cite{bulla_hewson_98}. This leads to the expression: 
\begin{equation}
\boldsymbol{\Sigma}_{\alpha \gamma} (i \omega_n) = \sum_{\gamma, \gamma'} \mathbf{F}_{\gamma \alpha \gamma'} \mathbf{G}_{\gamma' \gamma}^{-1} - z J_{\alpha} \sum_{\gamma'} \boldsymbol{\Phi}_{\alpha \gamma'} \mathbf{G}_{\gamma' \gamma}^{-1}
\end{equation}
with $ \boldsymbol{\Phi}_{\alpha \gamma} = \phi_\alpha^{\rm in} \phi_\gamma \left( \begin{smallmatrix} 1 & 1 \\ 1  & 1  \end{smallmatrix} \right) \delta_{n0}$ and
correlation functions $F^d_{\gamma \alpha \gamma} = G_{\hat b_\gamma^\dagger \hat b_\gamma \hat b_\alpha, \hat b_\gamma^\dagger}$ and $F^o_{\gamma \alpha \gamma} = G_{\hat b_\gamma^\dagger \hat b_\gamma \hat b_\alpha^\dagger, \hat b_\gamma^\dagger}$ [where we note $G_{\hat A, \hat B} (\tau) = \langle T_\tau \hat A(\tau) \hat B(0) \rangle$ and $T_{\tau}$ is the imaginary-time ordering operator], which are combined in the matrix $\mathbf{F}_{\gamma \alpha \gamma}$ according to $\mathbf{F} = \left( \begin{smallmatrix} F^d & F^o \\ (F^o)^* & (F^d)^* \end{smallmatrix} \right)$.

\section{Validity Issues }

\subsection{Validity Domain}

The BDMFT equations derived here rely on the limit of high dimensions, in particular implying a local self-energy. This means that phases in which the self-energy has a non-trivial dependence on momentum (like $p$-wave or $d$-wave pairing) cannot be captured. Whereas local correlations are taken into account exactly, non-local correlations are treated on the mean-field level.
Apart from these restrictions, BDMFT can be applied in a wide variety of settings; both the homogeneous case and the inhomogeneous case can be treated, the latter by a straightforward extension towards Real-Space BDFMT, analogous to Real-Space DMFT for fermionic atoms \cite{helmes_costi_08, snoek_titvinidze_08}. The scheme is completely flexible regarding the spin of the atoms involved \cite{li_bakhtiari_11}.
 
\subsection{Relevance to Other Theories}

The BMDFT equations as derived here incorporate Gutzwiller mean-field theory as the lowest order terms in $1/z$ \cite{rokhsar_kotliar_91, sheshadri_krishnamurthy_93}. The second order terms correspond to the dynamical mean-field terms familiar from fermionic DMFT \cite{georges_kotliar_96}.

\section{Application: Ttwo-Component Mixture on the Bethe Lattice}

\begin{figure}
\centerline{\includegraphics[width=10cm]{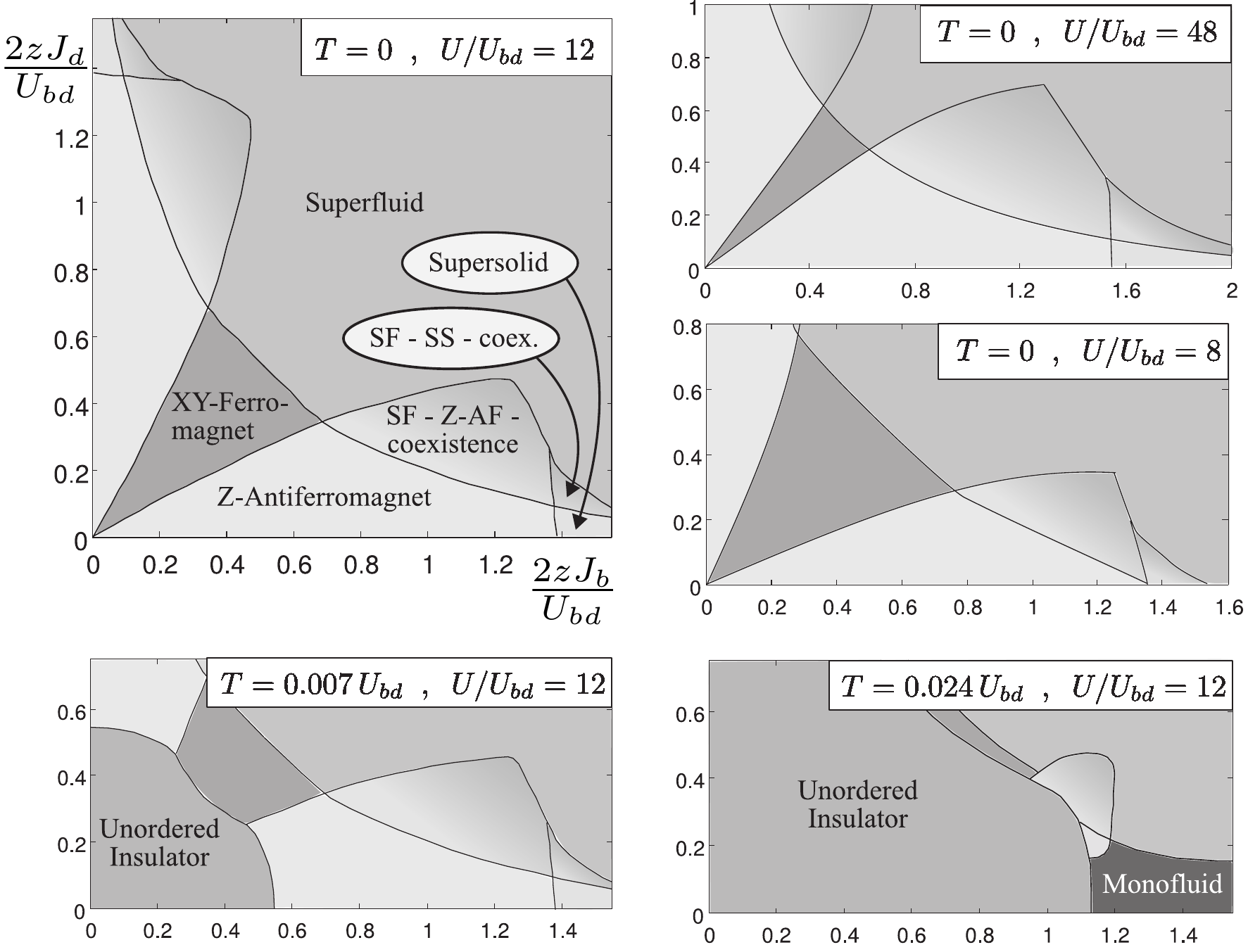}} 
\caption{Phase diagrams obtained by BDMFT for a Bose-Bose mixture on the Bethe lattice at total filling $n=1$ and $z=4$ \cite{hubener_snoek_09}.
} 
\label{snoek_fig1} 
\end{figure}

Applying BDMFT to a Bose--Bose mixture (denoted by $b$ and $d$) yields a rich phase diagram, as shown in Fig. \ref{snoek_fig1}. Here we have chosen symmetric interactions ($U_{bb}=U_{dd} = U$) and considered the case that $U_{bd} \ll U$. The total filling is fixed to one: $n_b+n_d=1$. The phase diagram includes a superfluid state and insulating states with spin order. The spin order is ferromagnetic in the $xy$-plane when the tunneling ampitudes $t_{b,d}$ for the two species are comparable and antiferromagnetic in the $z$-direction for a larger imbalance in effective mass \cite{altman_hofstetter_03, soyler_capogrosso_09, hubener_snoek_09}. The transition between those two 
types of spin ordered states is of first order.  
For very anisotropic tunneling amplitudes, we find a phase in which the species with the small tunneling amplitude becomes localized at every other lattice site, whereas the second species is still superfluid \cite{soyler_capogrosso_09, hubener_snoek_09, capogrosso_soyler_10}. Since in this phase off-diagonal superfluid order occurs together with spontaneous translational symmetry breaking, this is a supersolid phase \cite{hubener_snoek_09}.    

Above the critical temperature  for spin order (cf. the lowest two panels in Fig. \ref{snoek_fig1}) the magnetic order disappears, giving rise to an unordered insulator. The supersolid phase melts into a monofluid phase, in which one of the species is still superfluid, but translational symmetry is restored. 

\section*{Acknowledgments}
This work was supported by by the Nederlandse Organisatie voor Wetenschappelijk
Onderzoek (NWO) and the German Science Foundation DFG via Forschergruppe FOR 801 and Sonderforschungsbereich SFB/TR 49.

\newpage

\bibliographystyle{ieeetr}
\bibliography{BDMFT_chapter}

\end{document}